\documentclass{ws-ijmpd}

\usepackage{amsmath,amssymb}
\usepackage{epsfig}

\DeclareMathOperator{\extdm}{d}
\newcommand{\extd}{\extdm \!}

\newcommand{\beqas}{\begin{eqnarray*}}
\newcommand{\beqa}{\begin{eqnarray}}

\newcommand{\eeqas}{\end{eqnarray*}}
\newcommand{\eeqa}{\end{eqnarray}}

\newcommand{\eq}[2]{\begin{equation} #1 \label{#2} \end{equation}}

\newcommand{\al}{\alpha}
\newcommand{\be}{\beta}
\newcommand{\ga}{\gamma}
\newcommand{\de}{\delta}

\newcommand{\ka}{\kappa}
\newcommand{\la}{\lambda}

\newcommand{\De}{\Delta}
\newcommand{\Om}{\Omega}

\newcommand{\blist}{\begin{itemize}}

\newcommand{\elist}{\end{itemize}}

\newcommand{\Laa}{L}

\providecommand{\href}[2]{#2}

\DeclareFontFamily{OT1}{rsfs}{}
\DeclareFontShape{OT1}{rsfs}{m}{n}{ <-7> rsfs5 <7-10> rsfs7 <10->rsfs10}{} 
\DeclareMathAlphabet{\mycal}{OT1}{rsfs}{m}{n}

\newcommand{\V}{K}

\begin{document}

\markboth{H. Balasin and D. Grumiller}{Non-Newtonian behavior in weak field general relativity for extended rotating sources}

\newcommand{\oper}{\De}

%\title{General relativity reduces galactic dark matter by about 30\%} 

\title{NON-NEWTONIAN BEHAVIOR IN WEAK FIELD GENERAL RELATIVITY FOR EXTENDED ROTATING SOURCES}

%\title{Breakdown of the Newton approximation in galactic physics}

\author{\footnotesize H. BALASIN \footnote{hbalasin@tph.tuwien.ac.at}} 
\address{Institut f\"ur Theoretische Physik, Technische Universit\"at Wien,
Wiedner Hauptstr.~8-10/136, A-1040 Vienna, Austria}

\author{\footnotesize D. GRUMILLER\footnote{E-mail: grumil@lns.mit.edu}} 
\address{Massachusetts Institute of Technology, 77 Massachusetss Ave, Cambridge, MA 02139}

\maketitle

\begin{abstract}

Exact stationary axially symmetric solutions of the 4-dimensional Einstein equations with co-rotating pressureless perfect fluid sources are 
 studied. %This is of physical relevance for the dynamics of galaxies and questions concerning dark matter. 
A particular solution with approximately flat rotation curve is discussed in some detail. We find that simple Newtonian arguments over-estimate the amount of matter needed to explain these curves by more than $30\%$. The crucial insight gained by this model is that the Newtonian approximation breaks down in an extended rotating region, even though it is valid locally everywhere. No conflict with solar system tests arises.

\end{abstract}

%\pacs{04.20.Jb, 04.40.Nr, 95.35.+d, 98.62.Ck, 98.62.Gq}

%\preprint{TUW--05--2x, LU-ITP 2005/02x}

\maketitle

\section{Introduction}

The problem of the internal dynamics of galaxies is a very old one. It
is of special interest because observed non-Keplerian (henceforth
``flat'') rotation curves \cite{Rubin:1980zd} %,Bosma:1981} 
apparently
contradict the simple Newtonian results. % and therefore provide a strong evidence for dark matter, cf.~e.g.~\cite{Persic:1995ru}.  
%Two questions are pivotal here: {\em Do we have the correct theory of gravity?} Some attempts to avoid the dark matter hypothesis answer it negatively, for instance Modified Newton Dynamics \cite{Milgrom:1983ca}. So far these alternatives typically are at least as unsatisfactory as the dark matter hypothesis itself, cf.~e.g.~\cite{Zhytnikov:1994rm,Aguirre:2001xs}. The second question, {\em Are we applying our theory correctly?}, usually is answered in the affirmative. 
These observational facts are usually interpreted in favor of changing
the matter model, i.e.~introducing dark matter that accounts
for the difference between prediction and ``experiment'', cf.~e.g.~\cite{Persic:1995ru}. However, one
might also shift the burden  to the gravitational theory itself. The
most radical change in this line of arguments would result in avoiding
dark matter altogether, as for example in %the so-called 
MOdified Newtonian Dynamics \cite{Milgrom:1983ca}. %,Milgrom:1983pn}. 
So far these alternatives typically are at least as unsatisfactory as
the dark matter hypothesis itself,
cf.~e.g.~\cite{Zhytnikov:1994rm}. %,Aguirre:2001xs}. 

Taking a closer look at the usual approach, it is Newtonian Gravity rather than General Relativity (GR) that is applied to the situation of gravitational dynamics of galaxies. 
Because the velocity of the stars within a galaxy is small as compared
to the speed of light and since the gravitational field is very weak (at
least a couple of Schwarzschild radii away from a central galactic black
hole \cite{Eckart:1996}) %,Eckart:1997em}) 
the Newtonian approximation seems to be sufficient. However, one has to be careful with such arguments. For instance, it is clear that corrections from GR are smaller in regions where gravity is weaker -- and yet, the GR distortion of the elliptic orbit of Mercury visually is most pronounced in the aphelion, and not in the perihelion, although gravity is weaker in the former. It is the main purpose of this paper to show that indeed the Newtonian approximation breaks down, even though it might be valid locally everywhere in the Galaxy.

The idea to use GR rather than the Newtonian approximation 
%Quite recently there was an attempt to use GR rather than Newtonian physics 
to describe the internal dynamics of galaxies appears to go back to a
relatively recent paper by Cooperstock and Tieu
\cite{Cooperstock:2005qw}. %Ironically, they did not apply the theory
                           %correctly and were led to the conclusion
                           %that GR allows to avoid dark matter
                           %altogether. 
Several errors were pointed out in subsequent
publications, cf.~\cite{Korzynski:2005xq,Vogt:2005va} and \ref{app:B}. So it is fair to say that Ref.~\cite{Cooperstock:2005qw} caused a lot of confusion, interest and critical comments \cite{Korzynski:2005xq,Vogt:2005va,Garfinkle:2005qe}. %,Cooperstock:2005ba,Vogt:2005hv,Cross:2006rx,Rocek:2006bp,Bratek:2006uw,Fuchs:2006pr,Kostov:2006vp,Alba:2006ea,Lusanna:2006wx,Maia:2006ga,Zingg:2006mg}.
Actually, the approach of taking the non-linear character of GR
more serious fits quite well to similar considerations on a cosmological
scale \cite{Hosoya:2004nh}. %,Ellis:2005uz,Coley:2005ei,Wiltshire:2007jk}.

Having a simple framework for the construction of exact solutions %(rather than approximations or arguments) 
at our disposal helps to clarify these important issues. The prime goal of this paper is to provide and discuss such a framework and to point out where and why the Newtonian approximation breaks down.

%. As an application, a toy model for a galaxy is presented and the issue of flat rotation curves and dark matter is addressed. Finally, it will be shown that GR also predicts dark matter, albeit less than naive Newtonian estimates.

\section{Derivation and solution of field equations}

\subsection{Physical assumptions}
%Order of magnitude estimates of the typical electromagnetic fields and of the cosmological constant show that one can neglect such effects in the description of intragalactic dynamics at the sub-percent level. Thus, to leading order there are no relevant forces besides gravity. 
We do not discriminate between elliptic, spiral or irregular galaxies, but it is clear that some galaxies will obey the subsequent physical idealizations better than others.

Order of magnitude estimates reveal that there are no relevant forces besides gravity in the description of intragalactic dynamics at the percent level. 
The matter content of a galaxy can be modeled very well by a pressureless perfect fluid. The approximation of vanishing pressure is applicable in the whole galaxy, aside from the very center where a galactic black hole may be located and a region close to the rotation axis where jets might be emitted. 
Additionally, stationarity and axial symmetry will be imposed, which is reasonable for many galaxies. 
This implies the existence of two Killing vectors and considerably simplifies the geometry. 

We are thus interested in axially symmetric stationary solutions of the Einstein equations ($\ka=8\pi$, $G_N=c=1$)
\begin{equation}
  \label{eq:einstein}
  R^{ab}-\frac12 g^{ab}R = \ka T^{ab}\,,
\end{equation}
with the energy momentum tensor 
\begin{equation}
  \label{eq:ax2}
  T^{ab}=\rho u^a u^b\,,
\end{equation}
where $\rho,u^a$ are mass density and 4-velocity, respectively.

\subsection{Deriving the equations}
Theorem 7.1.1 of \cite{waldgeneral} (the premises of which apply to the case under consideration) guarantees that, without loss of generality, the line element can be brought into the following form: %(the speed of light is set to $c=1$):
\begin{equation}
  \label{eq:ax1}
  \extd s^2=-\V(\extd t-N\extd\phi)^2+\V^{-1}r^2\extd\phi^2+\Om^2(\extd r^2+\Laa\extd z^2)
\end{equation}
All functions $\V,N,\Om,\Laa$ depend solely on the non-Killing coordinates $r,z$. The coordinate $\phi$ is $2\pi$-periodic.
It should be noted that $\V=-\xi^a\xi_a$ is determined by the Killing norm of the timelike Killing vector $\xi^a=\partial_t^a$. We are not interested in inside solutions of black holes here, whence $\xi^a$ is timelike globally. %(close to a black hole horizon our assumption of vanishing pressure would break down anyhow, so in particular we have to assume that we are a couple of Schwarzschild radii away from a central galactic black hole).

We shall assume the perfect fluid velocity to be co-rotating, i.e., $u^a=(u^0(r,z), 0, 0, 0)$. So actually 
$u^a$ is proportional to the Killing vector $\xi^a=\partial_t^a$. The conservation equation $\nabla_a T^{ab}=0$ decomposes into
\begin{equation}
  \label{eq:ax3}
  \rho u^a\nabla_a u^b=0\,,\quad \nabla_a(\rho u^a)=0\,.
\end{equation}
If we suppose that $\rho\neq 0$ then the first equation implies geodesicity of the 4-velocity %-- and consequently geodesicity of the timelike Killing vector $\xi^a$ -- 
while the second one expresses stationarity of the mass distribution. 
%Now we make the pivotal observation that for the line element \eqref{eq:ax1} and $\rho \neq 0\neq u^0$ Eqs.~\eqref{eq:ax3} simplify to
%\begin{equation}
%  \label{eq:axnew1}
%  g^{ba}\partial_a g_{tt}=0\,,\quad \partial_t \rho = 0\,.
%\end{equation}
%Now we make the pivotal observation that also the timelike Killing vector $\xi^a$ is geodesic as a consequence of \eqref{eq:ax3}, 
This seemingly mild assumption entails that the Killing vector $\xi^a$ has to be geodetic and hence $\V$ is constant.
This is a crucial difference to the matterless case where no such restriction %on the Killing vector 
arises. 
%The standard manipulations
%\begin{equation}
%  \label{eq:ax123}
%  \xi^a \nabla_a\xi_b=-\xi^a\nabla_b\xi_a=-\frac{1}{2}\partial_b\left(\xi^a\xi_a\right)
%\end{equation}
%imply a constant Killing norm. 
%Consequently, the quantity $\V$ is a positive constant. 
Without loss of generality one may set $\V=1$ by a rescaling of the time coordinate in regions where $\rho\neq 0$. Henceforth the vacuum case will always be considered as the limit of small $\rho$. %, i.e., even in regions where $\rho\approx 0$ the property \eqref{eq:ax3} will be employed. 
Therefore, $\V=1$ is valid globally \footnote{While this eliminates pure vacuum solutions of technical interest it retains the physically important solutions where ``vacuum'' is just an approximation to ``very low energy density''.}. We emphasize that the presence of co-rotating pressureless perfect fluid sources {\em simplifies} the dynamics as compared to the vacuum case.

In addition one may set $\Laa=1$. The proof as given e.g.~in \cite{waldgeneral} is valid only for the vacuum case, but one can show easily that the crucial condition $R^t{}_t+R^\phi{}_\phi=0$ holds still for a pressureless perfect fluid (albeit it ceases to hold once pressure is switched on or a cosmological constant is added). Therefore, the line element \eqref{eq:ax1} reduces to
\begin{equation}
  \label{eq:ax4}
  \extd s^2=-(\extd t-N\extd\phi)^2+r^2\extd\phi^2+e^\nu(\extd r^2+\extd z^2)\,,
\end{equation}
with the redefinition $\Om^2=e^\nu$.
From now on we shall exclusively refer to this adapted coordinate system.
%This adapted coordinate system will be used exclusively from now on.
 
Inserting into the Einstein equations \eqref{eq:einstein}, \eqref{eq:ax2} establishes (cf.~\ref{app:A} and also \cite{vanStockum:1937,Stephani:2003tm})
\begin{align} 
& r\nu_z+N_rN_z=0\,,\label{eom1} \\ 
& 2r\nu_r+N_r^2-N_z^2=0\,,\label{eom2} \\
& \nu_{rr}+\nu_{zz}+\frac{1}{2r^2}\left(N_r^2+N_z^2\right)=0\,,\label{eom3}\\
& N_{rr}-\frac{1}{r}N_r+N_{zz}  = 0\,,\label{eom4}\\
& \frac{1}{r^2}\left(N_r^2+N_z^2\right)=\ka\rho e^{\nu}\,.\label{eom5}
\end{align}
%These are the exact Einstein equations. %without having invoked parturbation theory. 
Remarkably, the only difference to the perturbative results in \cite{Cooperstock:2005qw} is the factor $e^\nu$ on the right hand side in the last equation.

Before solving these equations we would like to point out that the 3-velocity as seen by an asymptotic observer who is at rest with respect to the center of the galaxy is
\begin{equation}
  \label{eq:velocity}
  V(r,z)=\frac{N(r,z)}{r} %\frac{N(r,z)}{\sqrt{r^2-N^2(r,z)}}\approx \frac{N(r,z)}{r}\,,
\end{equation}
provided $r>N(r,z)$. %$r^2\gg N^2(r,z)$.
This may be derived by a standard ADM split of the line-element
\begin{multline}
\extd s^2=-r^2/(r^2-N^2) dt^2 + (r^2-N^2)[d\phi + N/(r^2-N^2)dt]^2 %\\
+e^\nu(\extd r^2+\extd z^2)\,.
\label{eq:ADMsplit}
\end{multline}
The standard decomposition of $\partial_t^a$ into lapse and shift obtains
\eq{
\partial_t^a=\gamma (n^a + V E_\phi^a)\,,
}{eq:ADMsplit2}
where $n^a=E_t^a$ is the unit normal to the $t=\rm const.$ hypersurfaces and the shift vector is proportional to $E^a_\phi$, as expected for angular rotation. The relativistic $\ga$-factor is given by $\ga=1/\sqrt{1-V^2}$ with
\eqref{eq:velocity}.
So $V$ is the velocity distribution of the (co-)rotating dust as seen by an asymptotic observer who is at rest with respect to the rotation axis. The results above can be found in the classic textbook \cite{Stephani:2003tm} (cf.~also \cite{vanStockum:1937}), but we have reviewed them in some detail here in order to make explicit the relevance of each of our assumptions, and also because the result \eqref{eq:velocity} for the velocity is crucial for our discussion.

\subsection{Solving the equations}
The strategy for solving this coupled system of partial differential equations (PDEs) is as follows: first, we find the general solution of the linear second order PDE \eqref{eom4} for $N$ in terms of a spectral density. Then we use this solution to determine $\nu$ from \eqref{eom1}-\eqref{eom3} up to an integration constant. Finally, we solve the last equation \eqref{eom5} for $\rho$ with $N$ and $\nu$ as input. To match with input from observations we must fix the spectral density (and thus the function $N$) accordingly. From \eqref{eq:velocity} we see that $N$ is simply the velocity $V$ times the coordinate $r$. Thus, the strategy is to take $V$ as an input and to deduce the mass density $\rho$, rather than the other way around.

\subsubsection{Separation Ansatz and zero mode}
We implement now this strategy and focus in this Section on \eqref{eom4}.
The separation Ansatz $N={\cal R}(r){\cal Z}(z)$ yields ${\cal Z}_{zz}=k{\cal Z}$ with the separation constant $k\in\mathbb{R}$. Because of reflection symmetry $N(r,z)=N(r,-z)$. Therefore, ${\cal Z}(z)=\int\extd k A(k){\cal Z}(k,z)$ solely is composed of even modes ${\cal Z}(k,z)=\cosh{(\sqrt{k}z)}$. If $k>0$ then these modes diverge exponentially for $|z|\to\infty$. This is unphysical behavior. In \cite{Cooperstock:2005qw} a positive value for $k$ {\em and} fall off behavior at infinity was achieved by employing the non-smooth modes ${\cal Z}(k,z)=e^{-\sqrt{k}|z|}$, which satisfy \eqref{eom4} for $|z|>0$ only. Consequently, there are sources localized at $z=0$. We show in \ref{app:B} that either the weak energy condition is violated in the whole galactic plane or the pressure must be negative, so these sources are unphysical and contradict our assumption of vanishing pressure. Thus, we have no choice but to assume $k=-\la^2$ with $\la\in\mathbb{R}_0^+$ yielding the modes ${\cal Z}(\la,z)=\cos{(\la z)}$.

Let us address first the zero mode $\la=0$. Without loss of generality ${\cal Z}=1$ and ${\cal R}=A_0 + B_0 r^2$. Thus, the zero mode solution reads
\begin{equation}
  \label{eq:ax7}
  N_{0}=A_0+B_0 r^2\,.
\end{equation}
We consider first the simplest case, $B_0=0$. Then the velocity \eqref{eq:velocity} is given by $A_0/r$ and exceeds the speed of light for small $r$. While this is unpleasant, we note that the Einstein equations \eqref{eom1}-\eqref{eom5} are invariant under the constant shift $N\to N-A_0$. So the constant $A_0$ may be absorbed into a redefinition of the time variable. Thus, in addition to the fall-off condition for $|z|\to\infty$ exploited above and to the fall-off condition for $r\to\infty$ exploited below we have to impose a condition to fix $A_0$, e.g.~by requiring $N(0,0)=0$. This will be discussed in more detail below; right now we simply fix $A_0=0$. Then, if $B_0\neq 0$, the velocity profile  $V=B_0 r$ is linear, as expected from Newtonian gravity for a disk rotating with constant angular velocity. To get a first glimpse at possible differences between Newton and Einstein we consider the simplest case when only the zero mode is switched on, i.e., $V\propto r$, and restrict ourselves to the galactic plane $z=0$. In Newtonian gravity one obtains  
\begin{equation}
  \label{eq:ax20.3}
  \rho_N(r) = \rm const. 
\end{equation}
However, general relativity behaves differently: plugging $N\propto r^2$ into \eqref{eom2} yields $\nu\propto r^2$ with negative proportionality constant. The unperturbed Einstein equation \eqref{eom5} establishes ($\al\in\mathbb{R}$)
\begin{equation}
  \label{eq:ax20}
  \rho(r)\propto e^{\al^2r^2}\,.
\end{equation}
Conceptually (although not physically) this is an important result: given as input the same velocity profile in the galactic plane, $V\propto r$, Newtonian and general relativistic calculations yield different results for the mass density, even though locally the quantity $\rho$ as given in \eqref{eq:ax20} can always be matched to the Newtonian $\rho_N$ in \eqref{eq:ax20.3} by adjusting appropriately in $\nu$ the integration constant $\al$. Thus, only for a small region of spacetime the Newtonian picture is correct. However, the zero mode obviously is unphysical because for large values of $r$ the velocity $V$ would exceed the speed of light. Thus, from now on we set $B_0=0$, i.e., the zero mode is dropped. 

\subsubsection{Generic modes}
\begin{figure}
\centering
\epsfig{file=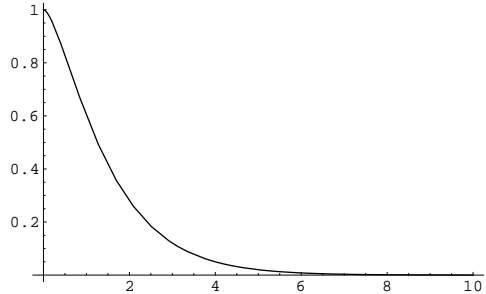,width=0.5\linewidth}
\caption{$x\cdot K_1(x)$}
\label{fig:1}
\end{figure}

In order to get the generic modes we use the rescaled variable $x=\la r$ and obtain from the separation Ansatz above the ordinary differential equation
\begin{equation}
  \label{eq:ax6}
  {\cal R}''-\frac{1}{x}{\cal R}^\prime-{\cal R}=0\,,
\end{equation}
where prime denotes differentiation with respect to $x$. It can be solved by standard methods. The result for $N={\cal RZ}$ is given by 
\begin{equation}
  \label{eq:ax8}
  N(r,z)=\int\limits_0^\infty\extd\la \cos{(\la z)}(r\la)\left[A(\la)K_1(\la r)+B(\la)I_1(\la r)\right]\,.
\end{equation}
The spectral densities $A(\la)$ and $B(\la)$ may be chosen according to the behavior required of $N(r,z)$. The functions $I_1$ and $K_1$ are modified Bessel functions of the first and second kind, respectively (the latter is also known as Macdonald function). We note that $I_1$ blows up exponentially for large values of $r$ which is unphysical. Therefore, we set $B(\la)=0$. The function $K_1$ falls off exponentially for large values of $r$ and diverges like $1/r$ near $r=0$. However, this divergence is compensated by a linear prefactor, so the integrand is well defined for any sufficiently regular $A(\la)$. We plot $x\cdot K_1(x)$ in Fig.~\ref{fig:1}.
The technical observation
\begin{equation}
  \label{eq:ax23}
  \int\limits_0^\infty \extd x x K_1(x) \cos{(ax/r)} = \frac{\pi}{2}\frac{r^3}{(a^2+r^2)^{3/2}}\,,
\end{equation}
will be very useful below.

To summarize, the general solution of \eqref{eom4} consistent with physical fall-off behavior $|z|,r\to\infty$ and with the absence of spurious sources in the galactic plane $z=0$ is given by
\eq{
N(r,z)=A_0+\int\limits_0^\infty\extd\la \cos{(\la z)}(r\la)A(\la)K_1(\la r)\,.
}{eq:ax30}
The constant $A_0$ has to be fixed by some additional boundary condition, e.g.~by requiring a certain behavior of $N$ as $r\to 0$. This can be done efficiently by setting first $A_0=0$, designing a useful spectral density $A(\la)$ and finally shifting $N$ by a suitable constant. 

\subsection{Fixing the spectral density}\label{se:2}

The choice of $A(\la)$ determines not only the radial velocity profile, but also the $z$-dependence of $N$. Thus, the behavior in $z$-direction is determined uniquely by the behavior in $r$-direction and vice versa. This is good for the predictability of the model, because one may fit one kind of behavior and predict the other, in addition to predicting the matter density. Instead of fixing $A(\la)$ directly we found it useful to perform first a Fourier transformation,\begin{equation}
  \label{eq:ax25}
  A(\la)=\frac{2}{\pi}\int\limits_0^\infty\extd x \,C(x)\cos{(\la x)}\,,
\end{equation}
where $A(\la)$ is determined in terms of a (Fourier) transformed spectral density $C(x)$. This allows to represent $V(r,0)$ as 
\begin{equation}
  \label{eq:ax24}
  V(r,0)=r \int\limits_0^\infty \extd x \frac{C(x)}{(x^2+r^2)^{3/2}} =\frac{1}{r} \int\limits_0^\infty \extd \tilde{x} \frac{C(\tilde{x}r)}{(\tilde{x}^2+1)^{3/2}}\,,
\end{equation}
with some arbitrary function $C(x)$, which we may call again 'spectral density'. It is evident that for $C\propto x^2$ the velocity profile is linear in $r$, while for $C\propto x$ the velocity profile is flat. Therefore, this purely technical step is very useful for matching with observational data. The function $N(r>0,z)$ reads 
\begin{subequations}
\label{eq:N}
\begin{multline}
  \label{eq:ax41}
  N(r>0,z) %=\frac{2}{\pi r}\int\limits_0^\infty\extd x C(x)\int\limits_0^\infty\extd\tilde{x}\cos{(\tilde{x}z/r)}\cos{(\tilde{x}x/r)}\tilde{x}K_1(\tilde{x})\\
= \frac{r^2}{2}\int\limits_0^\infty\extd x C(x)\Big(((z+x)^2+r^2)^{-3/2} %%\\
+((z-x)^2+r^2)^{-3/2}\Big)
\end{multline}
For $r=0$ we obtain
\begin{equation}
  \label{eq:ax41.5}
  N(0,z\geq 0)=\frac{2}{\pi}\int\limits_0^\infty\extd\la\cos{(\la z)}\int\limits_0^\infty\extd x C(x)\cos{(\la x)}
=C(z)\,,
\end{equation}
\end{subequations}
provided the function $C(z)$ is sufficiently well-behaved. The result for $z<0$ is obtained from reflection symmetry.

For physical applications one still has to choose the spectral density $C$ suitably.
It is instructive to consider the case of constant $C$: the function $N(r,z)$ turns out to be constant, which is problematic because for arbitrarily small values of $r$ the function $V(r,z)$ as defined in \eqref{eq:velocity} grows without bound. In model building we have encountered this feature to be generic, rather than an artifact of a particular choice of $C$: if the velocity profile is taken from experimental data it turns out that for some values of $z$ there will always be a region close to the axis $r=0$ where $V$ exceeds unity. This is a clear signal of the breakdown of our assumptions. For galaxies emitting jets this is not unexpected since jets emitted along the axis are not pressureless. Also the region around eventual central galactic black holes is located on the axis, so perhaps there is good (physical) reason for these regions not being described by a dust-like matter model. 
%these sources are not a ``bug'' but a ``feature''. 
It is not the purpose of the current paper to further discuss these regions, where closed time like curves may arise, similar to the G\"odel universe (the appearance of such regions appears to be a generic feature of axially symmetric stationary solutions).
Therefore, we cut out the domains where $N(r,z)\gtrsim r$ and blithely ignore them (cf.~Fig.~\ref{fig:patch}). While this is clearly unsatisfactory, it appears to be not much worse than ignoring the problem with cusps in Newtonian cold dark matter models.

\section{A toy model for galaxies}

\subsection{Input}

Experimental data imply \cite{Rubin:1980zd} %,Bosma:1981} 
that for small values of $r$ the velocity should increase linearly with $r$ and for intermediate values of $r$ it should yield an approximately flat profile. Asymptotically it should fall off like $1/r$ in order to match the Kerr solution \cite{Kerr:1963ud}. Moreover, it is required that $V(0,z)$ is regular for $|z|<r_0$, where the quantity $r_0$ is of the order of $1\,\,kpc\approx 3\cdot 10^{54}$, which allows to describe also the bulge region of the galaxy within our model.

With these remarks in mind we propose a simple 3-parameter function for $C$ ($\theta$ is the step function):
\begin{equation}
  \label{eq:ax43}
  C(x)=(x-r_0)(\theta(x-r_0)-\theta(x-R))+(R-r_0)\theta(x-R)
\end{equation}
The three parameters, $V_0,r_0,R$, have to be chosen appropriately. The quantity $V_0$ is of the order of the velocity in the flat regime ($V_0\approx 200\,\,km/s\approx 7*10^{-4}$). The bulge radius $r_0$ is set to $1\,\,kpc$ and $R$ to $100\,\,kpc$. 
Note that $C$ is smooth besides the points $x=r_0,R$. We stress that a smooth version of $C$ may be chosen to improve the model (and also other refinements may be implemented according to phenomenological necessities), but we want to keep the model simple and exactly soluble for pedagogic reasons, while still being realistic enough to be applicable to galactic physics.

\subsection{Output}

From \eqref{eq:ax43} and \eqref{eq:ax41} we obtain
\begin{multline}
  \label{eq:ax44}
  N(r,z)=V_0(R-r_0)+\frac{V_0}{2}\sum_\pm\left(\sqrt{(z\pm r_0)^2+r^2} %\right.\\\left.
-\sqrt{(z\pm R)^2+r^2}\right)\,,
\end{multline}
%The limit $r\to 0$ is described correctly as well. 
It may be checked that \eqref{eq:ax44} indeed satisfies \eqref{eom4} up to some distributional contributions at $r=0, |z|\geq r_0$. The latter lie in the region we are not capable to describe. It is an open problem how to deal with that region along the axis (even in the Newtonian approach), and we shall not address this issue here any further. However, as we shall see now, these critical regions lie outside the toy galaxy, i.e., at values $|z|\geq r_0$, and are localized along the $z$-axis. Therefore, we are able to draw meaningful conclusions for galactic rotation curves.

%For finite values of $r$ or for $|z|>R$ there are no sources besides the perfect fluid, as required. %In the limit of either $r$ or $|z|$ going to infinity, $N$ goes to the constant $V_0(R-r_0)$.
%\begin{figure}
%\epsfig{file=Nrz.eps,width=0.5\linewidth}
%\caption{\label{fig:galaxy}$N(r,z)$ exhibits how the galaxy curves spacetime}
%\end{figure}
%Fig.~\ref{fig:galaxy} shows the function $N(r,z)$ in the range $r\in(0,10R)$, $z\in(-10R,10R)$. One unit corresponds to $1\,\,kpc$ in that plot.
%One can actually ``see'' the bending of spacetime by the galaxy just by looking at this function. 

\begin{figure}
\epsfig{file=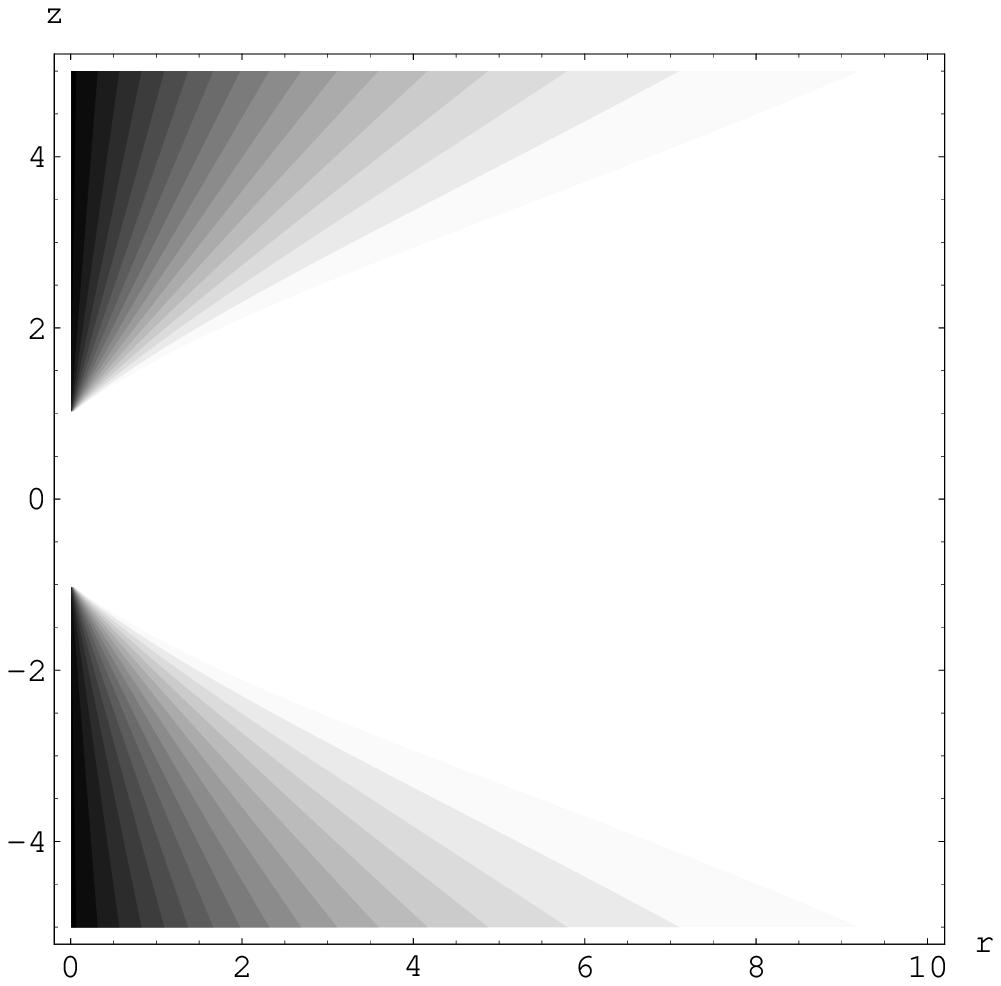,width=0.4\linewidth}
\epsfig{file=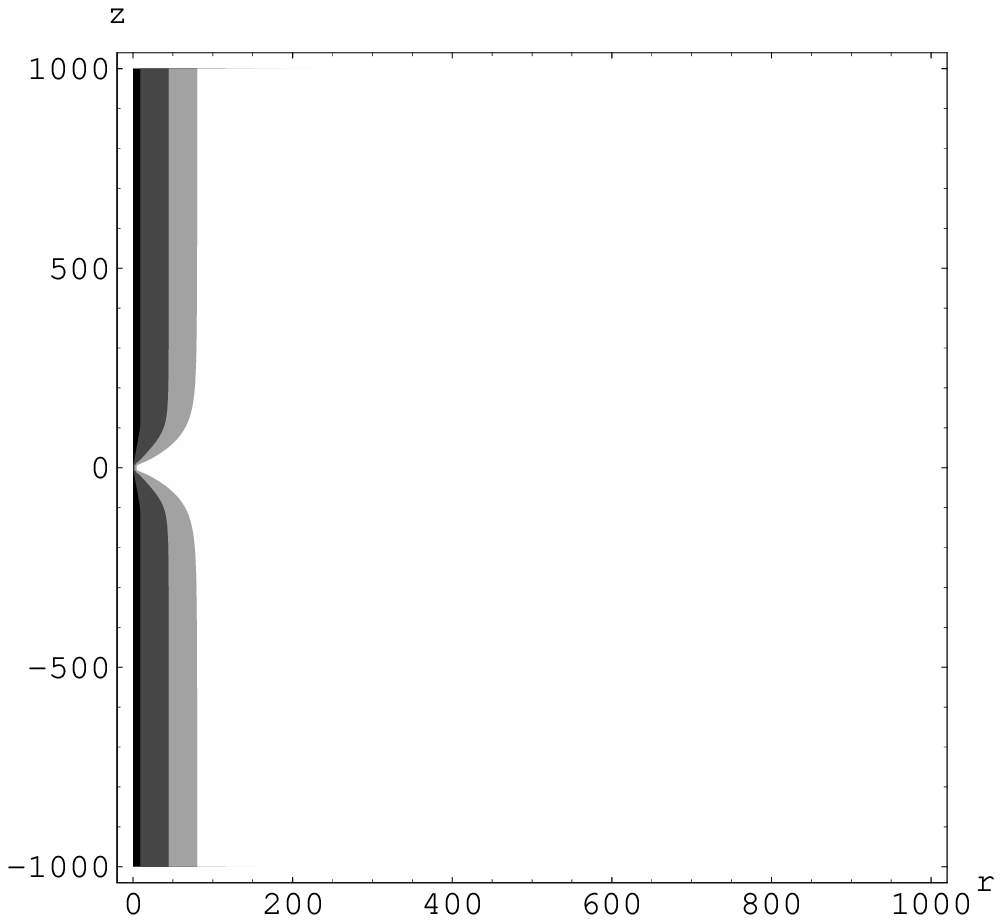,width=0.42\linewidth}
\caption{\label{fig:patch} Contour plot of $V(r,z)$ ($N(r,z)\gtrsim r$ in black region)}
\end{figure}

Contour plots of the velocity are depicted in Fig.~\ref{fig:patch}. The darker the region the higher the velocity. One unit corresponds to $1\,\,kpc$ in these plots. The left figure shows that indeed the function $V(0,z)$ is regular until $|z|=1\,\,kpc$, while the right figure reveals that on large scales the region we had to cut out stays close to the axis.
The velocity profile in the galactic disk reads
\begin{equation}
  \label{eq:ax46}
  V(r,0)=\frac{V_0}{r}\left(R-r_0+\sqrt{r_0^2+r^2}-\sqrt{R^2+r^2}\right)\,.
\end{equation}
Various limits are $V(r\ll r_0,0)\approx  r\,V_0/(2r_0)$, $V(r_0\ll r\ll R,0)\approx V_0$ and $V(R\ll r,0)\approx V_0 R/r$.
\begin{figure}
\epsfig{file=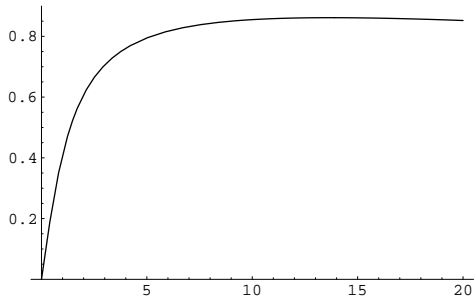,width=0.5\linewidth}
\caption{\label{fig:velo}$V(r,0)/V_0$ plotted up to $20\,\,kpc$}
\end{figure}
Figure \ref{fig:velo} depicts the ratio $V(r,0)/V_0$ up to $20\,\,kpc$. 
The velocity profile approximately behaves as required. %In particular, \eqref{eq:ax47a} reveals $V_0$ as the limiting velocity in the flat regime.
Remarkably, $r_0$ determines also the transition between the linear and the flat regime of the velocity profile, whereas the choice of $R$ determines the transition between the flat and the asymptotic regime. Thus, within this model the thickness of the bulge may be predicted from the velocity profile.
 
We have calculated the function $\nu$ by solving \eqref{eom1}-\eqref{eom3} and found that it is negligible up to the sub-permil level, besides the integration constant. 
This is mainly due to the fact that $V_0^2/4\approx 10^{-7}\ll 1$. Thus, for most practical purposes $\nu(r,z)\approx -\ln{\beta}=\rm const.$, where $\be$ is an integration constant.  
The density may now be derived from \eqref{eom5},
%\begin{figure}
%\epsfig{file=massdensity.eps,width=0.5\linewidth}
%\caption{\label{fig:density} Predicted mass density $\rho(r,z)$ on large scales}
%\end{figure}
\begin{equation}
  \label{eq:ax52} 
  \ka\rho(r,z) \approx \frac{\beta}{r^2}\left(N_r^2+N_z^2\right)\,.
\end{equation}
Integrating the density over the white region in the right Fig.~\ref{fig:patch} yields about $10^{11}$ solar masses for the total mass, which is not unreasonable.
The density within the galactic disk behaves as follows: %\footnote{The density for $z\neq 0$ may also be studied and it is not found to be in perfect agreement with measured galactic densities. This is a consequence of the simplicity of the choice \eqref{eq:ax43}.}:
%\begin{equation}
%  \label{eq:ax53}
%  \ka\rho(r,0) \approx \beta V_0^2\left(\frac{1}{\sqrt{r_0^2+r^2}}-\frac{1}{\sqrt{R^2+r^2}}\right)^2
%\end{equation}
for $r\to 0$ is goes to a constant ($\beta V_0^2/r_0^2$), for $r_0\ll r\ll R$ it falls off like $1/r^2$ and for $r\gg R$ it falls off like $1/r^6$. 
  
\subsection{Comparison with Newton}
We shall now compare the mass density for a given velocity profile as predicted by simple Newtonian considerations with the GR mass density calculated above. We shall restrict ourselves to the galactic plane for simplicity.

The Newtonian density for a given velocity profile may be derived naively from equating the kinetic energy per mass, $V^2/2$, to the potential energy per mass, $(4\pi\int\rho_N r_s^2\extd r_s)/r_s$. The quantity $r_s$ denotes the spherical radius $r_s^2=r^2+z^2$, which coincides with the axial one in the galactic plane.
Differentiation establishes
\begin{equation}
  \label{eq:ax54}
  \rho_N(r_s,0)=\rho_N(r,0) \propto \frac{V^2+2rVV'}{r^2}\,.
\end{equation}
The proportionality constant is known, but irrelevant as we are interested solely in the ratio $\rho/\rho_N$ and $\rho$ contains still an arbitrary multiplicative constant $\beta$ which has to be fixed.
%In the ratio between \eqref{eq:ax53} and \eqref{eq:ax54} (with \eqref{eq:ax46}) the only ambiguity comes from the integration constant $\beta$ in \eqref{eq:ax53}. We choose it such that in the region where luminous matter dominates, $r \ll r_0$, the Newtonian density coincides with the one predicted by GR, like in the standard approach.

%Surprisingly, in the flat region the correct density turns out to be only approximately $75\%$ of the one predicted by \eqref{eq:ax54}. Thus, if matter obeys the Einstein equations but Newtonian physics is used to explain the rotation curves then spurious matter (``dark matter'') is predicted. In fig.~\ref{fig:DM} the dark matter density mimicked in this way by GR is plotted (the radial units are normalized approximately in $kpc$, i.e., $r_0=1$, $R=100$).
%\begin{figure}
%\centering
%\epsfig{file=pDM.eps,width=0.5\linewidth}
%\caption{\label{fig:DM}Plot of $(1-\rho/\rho_N)$}
%\end{figure}
%Obviously, in the flat region $r_0\ll r\ll R$ there is a substantial discrepancy between simple Newtonian and exact GR predictions. 

In order to get a result which is robust with respect to slight changes in the spectral density $C$ we assume an {\em arbitrary} function $C$ with the following two properties: 1.~it leads to an approximately linear velocity profile for small $r$ (``small'' meaning, small as compared to $1\,\,kpc$, but not so small that one is at the scale of the galactic black hole, e.g.~around $1-100\,\,pc$), 2.~it leads to an approximately flat velocity profile for large $r$ (``large'' meaning, large as compared to sub-$kpc$ but still smaller than the whole galaxy, e.g.~around $1-20\,\,kpc$). For any given velocity profile $V(r,0)$ one may compare the Newtonian prediction with the one from GR \footnote{Obviously for the Keplerian profile $V\propto 1/\sqrt{r}$ the ratio \eqref{eq:ax55} becomes singular as a consequence of $\rho_N=0$.}: 
\begin{equation}
  \label{eq:ax55}
  \frac{\rho}{\rho_N}=\beta\left(1+\frac{r^2(V')^2}{V^2+2rVV'}\right)
\end{equation}
Clearly, at each point in the galaxy we may achieve perfect agreement between Newtonian and GR predictions by choosing $\beta$ appropriately. This reflects the local validity of the Newtonian approximation. However, it is important to realize that globally $\beta$ can only be chosen once.
%Motivated by the dominance of luminous matter in the central region 
Because usually one requires consistency between Newtonian and GR predictions in the region where luminous matter dominates, i.e., in the linear regime ($V\propto r$), let us now fix $\beta$ such that %in the linear regime ($V\propto r$) Newtonian and GR predictions coincide, i.e.,
\begin{equation}
  \label{eq:ax56}
  \left.\frac{\rho}{\rho_N}\right|_{\rm linear}=1=\beta\left(1+\frac{r^2}{3r^2}\right)=\beta\,\frac{4}{3}\,.
\end{equation}
%Therefore, $\beta=3/4$. 
Consequently, in the flat regime ($V=\rm const.$) we obtain
\begin{equation}
  \label{eq:ax57}
  \left.\frac{\rho}{\rho_N}\right|_{\rm flat}=\beta=\frac{3}{4}\,.
\end{equation}
Thus, requiring consistency between Newton and GR in the linear regime necessarily leads to a considerable deviation in the flat regime: Newtonian estimates require $133\%$ of the mass density as compared to GR calculations in order to get the same velocity profile. We reiterate the key observation: the integration constant $\be$ emerging in the GR solution can only be fixed once globally. 

We stress that these considerations are of relevance only for extended rotating sources. Thus, they play no role for solar system tests because there the main gravitational source is localized in the Sun.

\section{Conclusions} Using a simple model for a galaxy we could show that the Newtonian approximation over-estimates the amount of matter by about a third, whereas GR reduces the amount of dark matter needed to explain the flat rotation curves. This does not imply that exotic dark matter does not exist as insinuated by \cite{Cooperstock:2005qw}. After all, dark matter is a ``$500\%$ effect'' (meaning that there is about five times as much of dark matter in the universe as ordinary matter) and there are several independent measurements and arguments which suggest the existence of dark matter, e.g.~from microlensing \cite{Alcock:1995dm}. %But it will be rewarding to reconsider the problem of non-Keplerian rotation curves by improving the simple model presented in this letter, based upon the exact solution \eqref{eq:ax41}. %This may also help to understand the balance between dark and luminous mass pointed out recently \cite{}.
But as astrophysical measurements reach unprecedented accuracy the unexpected GR correction in the $30\%$ range predicted in our paper certainly will not be negligible. It is emphasized that this prediction has been essentially independent from the particular choice of the spectral density $C$.
%\acknowledgments

It would be interesting to check the robustness of our results by relaxing some of our assumptions -- in particular co-rotation -- or (some of) the symmetry requirements.

%On a more technical note, the toy model \eqref{eq:ax43} may be improved in several ways. In particular, it will be gratifying to pursue the patching as outlined in the paragraph ``Spectral density'' in order to understand the intricate relationship between galactic jets and galactic rotation curves. 

\section*{Acknowledgments}
We thank D.~Ahluwalia-Khalilova, P.~Aichelburg 
and D.V.~Vassilevich 
for discussion and R.~Meinel for correspondence. 
DG is grateful to T.~Buchert and D.~Schwarz, and L.~Bergamin for
 discussion in and invitation to Bielefeld and ESA, respectively. In addition DG would like to thank J.~Aman, I.~Bengtsson and N.~Pidokrajt for the hospitality at Stockholm University.
DG and HB acknowledge the repeated hospitality at the Vienna
University of Technology and at the University of Leipzig, respectively,
during the preparation of this paper. This paper is an extended version of the preprint {\tt astro-ph/0602519} by the present authors.

This work has been supported by project J2330-N08 of the Austrian
 Science Foundation (FWF), by project GR-3157/1-1 of the German Research
 Foundation (DFG) and during the final stage by project MC--OIF--021421
 of the European Commission and by project AO/1-5582/07/NL/CB, Ariadna ID 07/1301 of the European Space Agency.

\begin{appendix}

\section{Ricci tensor} \label{app:A}

In the adapted coordinate system \eqref{eq:ax4} the components of the Ricci tensor read
\begin{subequations}
\label{eq:Ricci}
\begin{align}
& R_{tt}=\frac{e^{-\nu}}{2r^2}\left(N_r^2+N_z^2\right)\\
& R_{t\phi}=\frac{e^{-\nu}}{2r^2}\left(-N(N_r^2+N_z^2)-r^2(N_{rr}-\frac{1}{r}N_r+N_{zz})\right)\\
& R_{tr}=0=R_{tz}\\
& R_{\phi\phi}=\frac{e^{-\nu}}{2r^2}\Big((r^2+N^2)(N_r^2+N_z^2) %\nonumber \\
%& \qquad\qquad 
+2r^2N(N_{rr}-\frac{1}{r}N_r+N_{zz})\Big)\\
& R_{\phi r}=0=R_{\phi z}\\
& R_{rr}=\frac{1}{2}\left(-\nu_{rr}-\nu_{zz}+\frac{1}{r^2}N_r^2+\frac{1}{r}\nu_r\right)\\
& R_{rz}=\frac{1}{2}\left(\frac 1r \nu_z+\frac{1}{r^2} N_rN_z\right)\\
& R_{zz}=\frac{1}{2}\left(-\nu_{rr}-\nu_{zz}+\frac{1}{r^2}N_z^2-\frac{1}{r}\nu_r\right)
\end{align}
\end{subequations}
As usual, on the right hand side indices $r$ and $z$ denote partial derivatives with respect to the corresponding coordinate and repeated indices denote repeated partial derivatives, e.g.~$N_{rr}:=\partial_r\partial_r N(r,z)$.

\section{Localized exotic energy-momentum tensor} \label{app:B}

In \cite{Cooperstock:2005qw} the expression ($J_1$ is a Bessel function of the first kind)
\eq{
N(r,z)=-\sum_n C(k_n) \sqrt{k}_n r J_1(\sqrt{k}_nr)e^{-\sqrt{k_n}|z|}
}{eq:CT1}
has been used, which is non-smooth at $z=0$. 
According to the
surface-layer formalism of Israel a localized energy-momentum contribution
is generated by the jump in the extrinsic curvature across the surface $z=0$,
\eq{
T_{ab}^{loc}=\ka\delta(z)([K_{ab}]-h_{ab}[K])
}{eq:CT2}
where $h_{ab}$ denotes (the continuous) induced metric and $[K_{ab}]=K_{ab}^{+}-K_{ab}^{-}$ the
jump of the extrinsic curvature. Inserting into the definition (${\cal L}_{n^e}$ denotes the Lie derivative along $n^c=E^c_z=e^{-\nu/2}\partial_{z}^c$)
\eq{
K_{ab}=\frac{1}{2}{\cal L}_{n^c}h_{ab} %h_{a}\,^{c}h_{b}\,^{d}
}{eq:CT3}
with $h_{ab}=-(\extd t-N\extd\phi)_{ab}^{2}+r^{2}\extd\phi_{ab}^{2}+e^\nu\extd r_{ab}^{2}$
we find
\eq{
K_{ab}=\frac{e^{-\nu/2}}{2}(2(\extd t-N\extd\phi)N_z\extd\phi+e^\nu\nu_z \extd r^{2})_{ab}\,.
}{eq:CT4}
The discontinuity $[N_z]=2\sum_n C(k_n)k_n rJ_1(\sqrt{k_n}r)$ inserted into \eqref{eom1} implies another discontinuity, $[\nu_z]=-N_r[N_z]/r$. The jump of the extrinsic curvature 
\eq{
[K_{ab}]=\frac{e^{-\nu/2}}{2}\left(2[N_z](\extd t-N\extd\phi)\extd\phi+e^\nu[\nu_z]\extd r^2\right)_{ab}
}{eq:CT5}
yields the localized (effectively 1+1 dimensional) energy-momentum tensor 
\eq{
T_{ab}^{\rm loc}=\ka\de(z)\frac{e^{-\nu/2}}{2r}N_r[N_z]\tilde{g}_{ab}:=P_I\tilde{g}_{ab}\,,
}{eq:CT6}
with the 1+1 dimensional metric
\eq{
\tilde{g}_{ab}=-(\extd t-N\extd\phi)^2_{ab}+r^2\extd\phi^2_{ab}+\frac{2r}{N_r}((\extd t-N\extd\phi)\extd\phi)_{ab}\,.
}{eq:CT9}
We note that both the 4- and 2-dimensional traces yield ${\rm Tr\,}T_{ab}^{\rm loc}=2P_I$, whereas both the 4- and 2-dimensional contractions with the time-like Killing vector $\xi^a$ yield the energy density 
\eq{
\rho_I:=T_{ab}^{\rm loc}\xi^a\xi^a=-P_I\,. 
}{eq:CT10}
It saturates the strong energy condition, and additionally obeys the weak energy condition if $N_r[N_z]\leq 0$ at $z=0$ for all $r$. We may interpret the induced energy-momentum tensor as the one of a 1+1 dimensional perfect fluid with mass density $\rho_I$ and pressure $P_I$. So either the pressure is negative or the weak energy condition is violated. In any case we have exotic matter\footnote{One can decide between these two possibilities by taking a certain velocity profile in the galactic plane, inserting into \eqref{eq:velocity}, calculating $N_r$ and checking its sign. For monomials $V=a^2 r^\al$ with $\al>-1$ we get $N_r>0$. In addition, one needs an argument for the sign of $[N_z]$.}. Moreover, one of our basic assumptions that led to the adapted line element \eqref{eq:ax4}, namely the absence of pressure, clearly is violated by \eqref{eq:CT6}.

So the Cooperstock-Tieu model is unphysical and inconsistent, as pointed out in numerous comments. Where applicable, results of this appendix essentially\footnote{We have the factor $\frac\ka2 e^{-\nu/2}$ in \eqref{eq:CT6} instead of $e^{-\nu}$ as in \cite{Vogt:2005va}, but this is a minor discrepancy and of no relevance for the argument that matter in the galactic plane is exotic in the Cooperstock-Tieu model.} agree with the ones of \cite{Korzynski:2005xq,Vogt:2005va}.

\end{appendix}

%\bibliographystyle{fullsort}
%\bibliography{review}

\providecommand{\href}[2]{#2}\begingroup\raggedright\endgroup

\end{document}